\documentclass[pra,twocolumn,a4paper,showpacs,superscriptaddress]{revtex4}
\usepackage{amsmath}
\usepackage{amsfonts}
\usepackage{graphicx}

\begin{document}

\title{Non-local properties of a symmetric two-qubit system}
\author{A. R. Usha Devi}
\email{arutth@rediffmail.com}
\author{M. S. Uma}
\author{ R. Prabhu}
\affiliation{Department of Physics, Bangalore University, Bangalore-560 
056, India}
\author{Sudha} 
\affiliation{Department of Physics, Thunga First Grade College, 
Thirthahalli-577432, India.}
\date{\today}

\begin{abstract}
Non-local properties of symmetric two-qubit states are quantified in terms 
of a complete set of entanglement invariants. We prove  that negative 
values of some of the invariants are signatures of  quantum entanglement.   
This leads us to identify  sufficient conditions for non-separability 
in terms of entanglement invariants. Non-local properties of two-qubit 
states extracted from (i) Dicke state (ii) state generated by one-axis 
twisting  Hamiltonian, and (iii) one-dimensional Ising chain with nearest 
neighbour interaction are analyzed in terms of the invariants 
characterizing them.   
\end{abstract}


\maketitle

\section{\label{sec:level1}Introduction}

Quantum entanglement is one of the most striking properties of quantum %%@
theory and it plays a crucial role in the  rapidly developing areas
such as quantum computation and quantum %%@
communication~\cite{Niel00,Bou00,Div95}, quantum teleportation~\cite{Ben93} %%@
and quantum cryptography~\cite{Ek91}. 
Entanglement reflects itself through non-local correlations among the %%@
subsystems of a quantum system. These {\it non-local properties} remain %%@
unaltered by local manipulations on the subsystems and provide a %%@
characterization of quantum entanglement. Two quantum states $\rho_1$ and %%@
$\rho_2$ are said to be {\em equally entangled} if they are related to each %%@
other through {\em local unitary operations}, which merely imply a choice %%@
of bases in the spaces of the subsystems. 
The non-local properties associated with a quantum state can be represented %%@
in terms of  a {\em complete set of local invariants}, which are functions %%@
involving 
the parameters of the quantum state, that remain unaffected by local %%@
unitary transformations. These {\it entanglement invariants} assume 
identical values for all equally entangled quantum states. 

Makhlin~\cite{Mak02} has studied the entanglement invariants of an %%@
arbitrary mixed state of two-qubits and proposed a complete set of 18 local %%@
invariants 
characterizing the system. A  set of 8 polynomial invariants has been %%@
identified in the case of pure three qubit states~\cite{Sud01}. Linden et. %%@
al.~\cite{Lin99} have outlined a  general prescription to identify the %%@
invariants associated with a multi particle system. In this paper 
we will focus on entanglement invariants of symmetric two-qubit systems, %%@
which  exhibit exchange symmetry. Such two-qubit systems (N=2) restrict %%@
themselves to a 
space spanned by the symmetric angular momentum states $|S, M\rangle$ with %%@
$S=\frac{N}{2}=1,\ M=-1, 0, 1$. 
We have organized this paper as follows: In Sec.~\ref{sec:level1} we %%@
identify that  the number of invariants required to characterize a 
symmetric two-qubit system reduces from  18 (proposed by %%@
Makhlin~\cite{Mak02}) to 6. Moreover, we consider a specific case of 
symmetric two-qubit system, which is realized in  several physically %%@
interesting examples like, even and odd spin states~\cite{Xi03}, 
Kitagawa - Ueda state generated by one-axis twisting %%@
Hamiltonian~\cite{Kit93},  
qubits in a one-dimensional Ising chain~\cite{Xi01},  steady state of %%@
two-level atoms in a squeezed bath~\cite{GSA90} etc. For this special case %%@
of 
symmetric two-qubit  system, we show that  a subset of three  
independent invariants is sufficient to characterize the non-local %%@
properties completely. Sec.~\ref{sec:level2} discusses
 separable  symmetric two-qubit states, for which  some of the  invariants %%@
necessarily assume positive values. 
Using this result,  we  propose  criteria,  which provide a %%@
characterization of non-separability (entanglement) in  symmetric two-qubit %%@
states. 
In Sec.~\ref{sec:level3} we discuss  physical examples of symmetric %%@
two-qubits 
 extracted from $N$ qubit systems like  (i) Dicke state, (ii) state %%@
generated by one-axis twisting 
 Hamiltonian~\cite{Kit93} and (iii) state extracted from a one-dimensional %%@
Ising chain with nearest neighbour interaction~\cite{Xi01}. 
Sec.~\ref{sec:level4} contains  a brief summary of our results. 

\section{\label{sec:level1} Entanglement Invariants}

\label{ent}

Density matrix of an arbitrary two-qubit state in the Hilbert-Schmidt space %%@
${\cal H}={\cal C}^2\otimes{\cal C}^2$ is given by 
\begin{equation}
\rho=\frac{1}{4}\left(I\otimes I+ \vec{s}\cdot (\vec{\sigma}\otimes %%@
I)+(I\otimes\vec{\sigma})\cdot \vec{r}
     +\sum_{i,j=1}^3 t_{ij}\,(\sigma_{i}\otimes\sigma_{j})\right ) \, , 
\label{rho} \end{equation}
where $I$ denotes the $2\times 2$ unit matrix; $\sigma_i,\ i=1,2,3$ are the %%@
standard Pauli matrices; $\vec{s}={\rm Tr}\,[\rho\, (\vec{\sigma}\otimes %%@
I)]$ and 
$\vec{r}={\rm Tr}\, [\rho\, (I \otimes\vec{\sigma})]$ denote average spins %%@
of the qubits; $t_{ij}={\rm Tr}\,[\rho\, (\sigma_{i}\otimes\sigma_{j})]$ %%@
are 
the elements of the 
real $3\times 3$  matrix  $T$ corresponding to two-qubit correlations. 
The set of 15 state parameters $s_i,\ r_i, \  t_{ij}$ transform under local %%@
unitary 
operations $U_1\otimes U_2$ on the qubits as follows: 
\begin{eqnarray}
\label{tran} 
&s'_i=\displaystyle\sum_{j=1}^3 O^{(1)}_{ij}s_j \, ,  \ \  %%@
r'_i=\sum_{j=1}^3  O^{(2)}_{ij}r_j \nonumber\\ & %%@
t'_{ij}=\displaystyle\sum_{k,l=1}^3 O^{(1)}_{ik}\, O_{jl}^{(2)}t_{kl}    %%@
{\rm \ \ \ or \ \ \ } \  
    T'=O^{(1)}\, T\,  O^{(2)\, \dag}\, ,  
\end{eqnarray}     
where $O^{(i)}\in SO(3,R)$ are the $3\times 3$ rotation matrices, uniquely %%@
corresponding to the $2\times 2$ unitary matrices $U_i~\in~SU(2).$ Quantum 
entanglement of a given two-qubit state remains unaltered under local %%@
unitary transformations $U_{1}\otimes U_2$ and has been quantitatively %%@
captured 
in terms of a complete set of 18 invariants~\cite{Mak02} given below: 
\begin{eqnarray} 
\label{inv}
I_1&=&{\rm det}\, T\, ,\ \   I_2={\rm Tr}\, (T^\dag\, T),\ \ 
I_3={\rm Tr} \ (T^\dag\, T)^2,\,\nonumber\\ 
I_4&=&s^\dag\, s\, , \ \ \ \ I_5=s^\dag\, T\, T^\dag\, s\, , \ \  \ %%@
I_6=s^\dag\ (T\, T^\dag)^2\,  s,  \nonumber\\ 
I_7&=&r^\dag\, r \, ,\ \ \ \ I_8=r^\dag\, T\, T^\dag\, r \, ,\ \ \   %%@
I_9=r^\dag\, (T\, T^\dag)^2\,  r \, ,\nonumber\\   
I_{10}&=&\epsilon_{ijk}\, s_i\, (T\, T^\dag\, s)_j\, ([T\,T^\dag]^2\,s)_k %%@
\, ,\hskip .5in \nonumber\\   
I_{11}&=&\epsilon_{ijk}\, r_i\, (T^\dag\, T\, r)_j\, ([T^\dag\,T]^2\,r)_k  %%@
\, ,\hskip .5in  \nonumber\\
I_{12}&=&s^\dag\,T\, r\ ,  \hskip 1in \ I_{13}=s^\dag\,T\,T^\dag\,T\, r \, %%@
, \hskip .5in\nonumber\\   
I_{14}&=& \epsilon_{ijk}\,\epsilon_{lmn}\, s_i\,r_l\, t_{jm}\, t_{kn} \, %%@
,\hskip .8in \nonumber\\   
 I_{15}&=&\epsilon_{ijk}\, s_i\, (T\, T^\dag s)_j\, (T\,r)_k \, ,\hskip %%@
.8in\nonumber\\   I_{16}&=&\epsilon_{ijk}\, (T^\dag\, s)_i\, r_j\,  %%@
(T^\dag\, T\, r)_k\,  \, , \hskip .6in\nonumber\\ 
I_{17}&=&\epsilon_{ijk}\, (T^\dag \, s)_i\, (T^\dag\, T\, T^\dag\, s)_j\, %%@
r_k \, ,  \hskip .5in\nonumber\\
I_{18}&=&\epsilon_{ijk}\, s_i\, (T\, r)_j\, (T\, T^\dag\, T\, r)_k  %%@
\,.\hskip .6in 
\end{eqnarray} 
It is easy to verify that all the $I_k,\ k=1,2,\ldots ,18,\ $ are %%@
invariant, when the state parameters  $s_i,\ r_i, \  t_{ij}$ 
transform - under local 
unitary transformations of the two-qubit state $\rho$ - as shown in %%@
~Eq.~(\ref{tran}).  Makhlin~\cite{Mak02} has given  an  explicit procedure 
to find local unitary operations that transform any equivalent density %%@
matrices to a specific form,  uniquely determined by the set of 
18 invariants given in ~Eq.~(\ref{inv}).  For example,  $I_1 - I_3$ %%@
determine the diagonal 
form $T^d$ of the correlation matrix $T$ (achieved through local operation %%@
$T^d=O^{(1)}\, T\, O^{(2)\dag}$); the invariants $I_4-I_6$ and $I_7-I_9$ 
specify the absolute values of the six components of 
the average spins $\vec{s}$ and $\vec{r}$ of the qubits; other invariants %%@
$I_{10}-I_{18}$ determine the signs of these 
components. It is thus shown~\cite{Mak02} that  the two-qubit states   %%@
$\rho_1$ and $\rho_2$ are locally equivalent iff 
the invariants $I_1-I_{18}$ are identical for these states. Further, any %%@
quantitative measure of entanglement should be a function of these %%@
invariants. 

We turn our focus to symmetric two-qubit systems, which respect exchange %%@
symmetry viz., 
$$\Pi_{12}\, \rho^{(\rm sym)}\, \Pi_{12}^{-1}~=~ \rho^{(\rm sym)},$$ 
$\Pi_{12}$ being the permutation operator. Quantum states of symmetric two
qubit $N=2$ systems get confined to a three dimensional subspace $\left\{\ %%@
\vert  \frac{N}{2},\ M\rangle ;M=\pm 1,0 \right\}$  of the Hilbert space %%@
${\cal C}^2\otimes{\cal C}^2$.
The state parameters of a symmetric two-qubit system satisfy the following   
constraints due to exchange symmetry~\cite{footnote1} : 
\begin{equation}
   \label{con}
 \hskip 0.6in  r_i= s_i\,  ,\ \  t_{ij}=t_{ji}\, ,\ \ {\rm Tr}\, (T)=1 \, , 
\end{equation} 
and  thus  8 real state parameters viz.,  $r_i$ and the elements $t_{ij}$ %%@
of the real symmetric correlation matrix $T$,  
which has unit trace, characterize a symmetric two-qubit system. 
For a two-qubit state obeying exchange symmetry it is easy to see that  %%@
$I_4~=~I_7,\ I_5~=~I_8,\ I_6~=~I_9,$ 
$I_{10}~=~I_{11},$  $I_{15}~=~I_{16}$ and $I_{17}~=~I_{18},$  
reducing the number of invariants to 9.  Moreover, since the  matrix $T$ is %%@
real symmetric with  ${\rm Tr} (T)=1$, the invariants 
$I_1$ and $I_2$ alone suffice to determine the diagonal form of the %%@
two-qubit correlation matrix $T$ ( i.e., $I_1=t_1\, t_2\, t_3\,$ and %%@
$I_2=t_1^2+t_2^2+t_3^2$ determine the 
eigenvalues $t_1,\  t_2,\  t_3$ of the real symmetric two-qubit correlation %%@
matrix $T$ in the case of symmetric qubits). In other words, 
for a symmetric two-qubit system, the subset $(I_1,\ I_2,\  I_3)$ of three %%@
invariants reduces to $(I_1,\ I_2)$.  
Along with $I_1,\ I_2$, the invariants 
\begin{eqnarray}
I_4&=&s_1^2+s_2^2+s_3^2, \nonumber\\
I_{12}&=&s_1^2\,t_1+s_2^2\, t_2+s_3^2\, t_3, \nonumber \\
I_{14}&=&2\,(s_1^2\,t_2\,t_3+s_2^2\, t_1\, t_3+s_3^2\, t_1\, t_2),
\end{eqnarray} 
allow for the determination of  $s_1^2, \ s_2^2$ and $s_3^2,$ thus fixing %%@
the absolute values of the components of qubit orientation vector 
$\vec{s}$. One more invariant,
\begin{equation}
I_{10}=\left(t_1^4\,(t_3^2-t_2^2)+t_2^4\,(t_1^2-t_3^2)+t_3^4\,(t_2^2-t_1^2)%%@
\right)s_1\,s_2\,s_3 
\end{equation} 
 fixes the relative signs of $s_1^2, \ s_2^2$ and $s_3^2$~\cite{footnote2}. 
Therefore {\em the subset $(I_1,\ I_2,\ I_4,\ I_{10},\ I_{12}\ {\rm  \ and\ %%@
} I_{14})$ of six invariants gives a complete characterization of the 
non-local properties of symmetric two-qubit states}.  

Many physically interesting cases of symmetric two-qubit states like for %%@
e.g.,  even and odd spin states~\cite{Xi03}, 
Kitagawa - Ueda state generated by one-axis twisting %%@
Hamiltonian~\cite{Kit93},  
qubits in a one-dimensional Ising chain~\cite{Xi01},  steady state of %%@
two-level atoms in a squeezed bath~\cite{GSA90},  exhibit a 
 particularly simple structure 

 \begin{equation}
 \label{cansym}
 \varrho^{(\rm sym)}=\left(\begin{array}{llll}
                     a & 0 & 0& b\cr 
                     0& c & c& 0\cr 
                     0& c & c& 0 \cr 
                     b^*& 0 & 0& d\cr 
              \end{array}\right) \, , 
\end{equation}
of the density matrix ( in the standard two-qubit basis $|0_1\, %%@
0_2\rangle\, , |0_1\, 1_2\rangle\, , |1_1\, 0_2\rangle\, , |1_1\, %%@
1_2\rangle$) and it will be 
interesting to analyze the non-local properties of such systems through %%@
local invariants. 
The  specific  structure  $\varrho ^{(\rm sym)}$ given %%@
by~Eq.~(\ref{cansym}) of the  two-qubit density matrix   further reduces %%@
the number of 
parameters essential for the problem. Entanglement invariants associated %%@
with the symmetric two-qubit system,  with 
a simple structure $\varrho^{(\rm sym)}$ given by~Eq.~(\ref{cansym})  for %%@
the state, may  now be identified through a 
simple calculation to be 

\begin{eqnarray} 
   \label{invsym}
 I_1&=&(4\, c^2-4\,|b|^2)\, (1-4\, c), \nonumber \\ 
 I_2&=&(2\, c+2\, |b|)^2+(2\, c-2\, |b|)^2+(1-4\, c)^2, \nonumber \\ 
 I_4&=&(a-d)^2\, ,\ \ \ \   I_{10}=0,\, \nonumber \\
 I_{12}&=&(a-d)^2\, (1-4\, c)\, ,  I_{14} =8\, (a-d)^2\, (c^2-\, |b|^2). %%@
\nonumber \\
& & 
\end{eqnarray}  
In this special case we can express the invariants $I_1$ and $I_2$  in %%@
terms of $I_4,\ I_{12} {\rm \ and\ } I_{14}\ \  ({\rm provided}\   I_4\neq %%@
0)$ :  
\begin{eqnarray}
I_1=\frac{I_{14}\, I_{12}}{2\ I_4^2},  &\ \ \ \ \ \ %%@
I_2=\displaystyle{\frac{(I_4-I_{12})^2-I_4\, I_{14}+I_{12}^2}{I_4^2}.}
\end{eqnarray}
If $I_4=0$, the set containing six invariants  reduces to  the subset of %%@
non-zero invariants   $(I_1,\  I_2)$ leading to the 
identification that {\em  the set $(I_4,\  I_{12},\  I_{14})$ 
{\rm ( or} $(I_1,\ I_2)$ {\rm if} $I_4=0)$ characterizes the non-local %%@
properties 
of symmetric two-qubit states having  a specific structure $\varrho^{\rm %%@
(sym)}$ given by~Eq.~(\ref{cansym}) for the density matrix}. 

\section{\label{sec:level2} Characterization of separable symmetric %%@
two-qubit states}
A separable symmetric two-qubit density matrix is an arbitrary convex %%@
combination of direct product of identical single qubit states
$\rho_w=\frac{1}{2}\left(I+\displaystyle\sum_{i=1}^3\sigma_i\, %%@
s_{wi}\right)$ and is given by 
\begin{equation}
\label{sep}
\rho^{\rm (sym-sep)}=\sum_w\, p_{w}\, \rho_w\otimes\rho_w ;\ \ \sum_w\, %%@
p_w=1\, . 
\end{equation} 
Separable symmetric system is a classically correlated system, which can be %%@
prepared through classical communications between two parties. 
One of the important goals of quantum information theory has been to %%@
identify and characterize separability. We look for such identifying %%@
criteria for separability, in terms of entanglement invariants, in the %%@
following theorem:

\noindent{\bf Theorem:} The invariants,  $ I_{14},\ I_{12}$ and a  %%@
combination  $I_{12}-I_4^2$  of the invariants, necessarily assume 
positive values  for a symmetric separable two-qubit state,  which has a %%@
non-zero value for the invariant 
$I_{4}.$ 

\noindent{\bf Proof:} First we note that in  a  separable symmetric system  %%@
the correlation matrix elements are 
given by $t_{ij}=\displaystyle\sum_w\, p_w\ s_{wi}\, s_{wj}\,, \ $ and the %%@
components of the average spin  of the qubits are given by 
$s_i=\displaystyle\sum_w\, p_w\ s_{wi}.$ We therefore obtain, 
\begin{eqnarray} 
\label{sep2} 
 (i)\ \ \ \ \ \ \ \   I_{12}&=&s^\dag\, T\, s=\displaystyle\sum_{i,j=1}^3\, %%@
t_{ij}\,s_i\, s_j \nonumber\\
&=&\displaystyle\sum_w\, p_w\,\left(\displaystyle\sum_{i=1}^3\, s_{wi}\, %%@
s_{i}\right)
\left(\displaystyle\sum_{j=1}^3\, s_{wj}\, s_{j}\right)\nonumber\\ 
&=&\displaystyle\sum_w\, p_w\,\left(\vec{s}_{w}\cdot \vec{s}\right)^2  \geq %%@
0 ; \nonumber\\ & \nonumber\\
(ii)\ \ \ \ \ \ \ \ I_{14}&=&\epsilon_{ijk}\,\epsilon_{lmn}\, s_i\,s_l\, %%@
t_{jm}\, t_{kn} \nonumber\\  
&=& \displaystyle\sum_{w,w'}\, p_{w}\, %%@
p_{w'}\left(\vec{s}\cdot(\vec{s}_w\times\vec{s}_{w'})\right)^2 \geq 0 %%@
;\nonumber\\
(iii)\ \ I_{12}-I_4^2&=&\displaystyle\sum_w\, p_w\, %%@
(\vec{s}_w\cdot\vec{s})^2\ -\left(\displaystyle\sum_{w}\, p_{w}\,  
(\vec{s}_w\cdot\vec{s}) \right)^2 \nonumber\\  
&=&\langle  (\vec{s}_w\cdot\vec{s})^2\rangle %%@
-\langle(\vec{s}_w\cdot\vec{s})\rangle ^2  \geq 0\, , 
\end{eqnarray} 
for a separable symmetric two-qubit system, proving the above theorem. 

We conclude that non-positive  values of    $I_{12},\  I_{14}$ or  %%@
$I_{12}-I_4^2$  serve  as  a signature of entanglement and hence 
provide  sufficient conditions for non-separability of the quantum state.
It would be interesting to explore how these constraints on the invariants, 
get related to the other well established criteria of entanglement.
For two qubits states, it is well known that Peres's PPT (positivity of %%@
partial transpose) criterion~\cite{peres} is both necessary and sufficient %%@
for separability. We now proceed to show that, in the case of symmetric %%@
states given by ~Eq.~(\ref{cansym}), there exists a simple connection %%@
between the Peres's PPT criterion and the non-separabiltiy constraints (see %%@
~Eq.~(\ref{sep2})) on the invariants.\\
It is easy to identify the eigenvalues of the partially transposed density %%@
matrix $\varrho^{\rm (sym)}$ of ~Eq.~(\ref{cansym}):
\begin{eqnarray*} 
 \lambda_{1,2}&=&\frac{1}{2}\left((a+d)\mp\sqrt{(a-d)^2+4c^2}\right)\\
   \lambda_{3,4}&=&c\,\mp\vert b\vert
\end{eqnarray*}
of which $\lambda_1$ and $\lambda_3$ can assume negative values. Now,
(i) for $\lambda_1<0$, we must have $(a+d)^2<(a-d)^2+4c^2$, which on using %%@
${\rm Tr} (\varrho^{\rm (sym)})=a+d+2c=1$, reduces to the inequality %%@
$(1-4c)<(a-d)^2$.
Obviously (see ~Eq.~(\ref{invsym})),
\begin{equation}
 I_{12}-I_4^2=(a-d)^2\left((1-4c)-(a-d)^2\right),
\end{equation}
associated with the state $\varrho^{\rm (sym)}$, is negative when %%@
$\lambda_1<0$.
Further, (ii) we may express the invariants $I_{14}$ (using %%@
~Eq.~(\ref{invsym})) corresponding to the state $\varrho^{\rm (sym)},$
in terms of the eigenvalue $\lambda_3$ as
\begin{eqnarray} 
\hskip 1.0in I_{14}=8\,(a-d)^2\,(c+\vert b\vert)\,\lambda_3,
\end{eqnarray}
from which it follows immediately that, $I_{14}<0$ if $\lambda_3<0$. 
We have thus established an equivalence between the PPT criterion and the %%@
constraints on the invariants for two qubit symmetric state of the form %%@
$\varrho^{\rm (sym)}.$
In other words, the nonseparability conditions $I_{12}-I_4^2,\, I_{14}<0,$ %%@
are 
both necessary and sufficient for a class of two-qubit symmetric states %%@
given by ~Eq.~(\ref{cansym}). However, we leave open the question whether %%@
nonseparability constraints $I_{12}-I_4^2,\,I_{12}<0,\, I_{14}<0$ on the %%@
invariants, serve as both necessary and sufficient for an arbitrary
symmetric two qubit system.

\section{\label{sec:level3} Examples of symmetric two-qubit states}
We consider some physical examples of two-qubit states, where random pairs %%@
of qubits are extracted from a symmetric state of $N$ qubits.

\subsection{Two-qubit system  drawn from a Dicke State}
Density matrix of a random pair of qubits picked from a symmetric Dicke %%@
state~\cite{Xi02}  $|S=\frac{N}{2}\, , M\rangle,\ 
-\frac{N}{2}\leq M\leq\frac{N}{2}, $ of $N$ qubits  has the simple  %%@
structure given by 
~Eq.~(\ref{cansym}) with 
\begin{equation}
a=\frac{(N+2M)\, (N+2M-2)}{4\, N\, (N-1)},\ \ \ b=0, \ \ \
c=\frac{N^2-4\, M^2}{4\, N\, (N-1)} \, 
\end{equation}\\
 and $d=1-a-2c.$ The invariants $I_4,\ I_{12},\ I_{14}\ {\rm and} \ %%@
I_{12}-I_{4}^2$ associated with this system are given by 
\begin{eqnarray} 
\label{invdicke}
 I_{4}=\left(\frac{2M}{N}\right)^2,  I_{12}=I_{4}\,\frac{(4\, %%@
M^2-N)}{N\,(N-1)},\nonumber \\ 
I_{14}=8\, I_{4}\, \left(\frac{N^2-4\,M^2}{4\, N\, (N-1)}\right)^2\nonumber %%@
\\ 
I_{12}-I_{4}^2=I_4\left(\frac{4M^2-N^2}{N^2(N-1)}\right)
\end{eqnarray} 
and it is easy to identify  that $I_{14}\geq 0,$ but  $I_{12}-I_4^2$ is %%@
negative, 
implying that the system is indeed non-separable (except for the states %%@
with $M=\pm\frac{N}{2})$. However it is interesting to identify that as %%@
$N\rightarrow\infty$, $I_{12}-I_4^2$ approaches zero, revealing that the %%@
system tends to be separable in this limit. 

\subsection{Two-qubit state extracted from non-linear one-axis twisting %%@
Hamiltonian}
Kitagawa and Ueda~\cite{Kit93} had  proposed a non-linear Hamiltonian %%@
$H=\chi\, S_x^2$, where $S_x$ is 
the $x$-component of collective angular momentum of a $N$ two level system, %%@
to generate multi-atom 
spin squeezed states. A random pair of two-qubits of such a spin squeezed %%@
state has  the density operator~\cite{Xi02} $\ \varrho^{\rm (sym)}$   with 
\begin{equation} 
\begin{array}{l}
a=\frac{1}{8}\, (3+\cos^{(N-2)}(2\,\chi\ t) -4\, \cos^{(N-1)} (\chi\, t), %%@
\\
{\rm Im}\,b=\frac{1}{2}\, \cos^{(N-1)}(\chi\, t)\sin (\chi\, t),\ \\
c=\frac{1}{8}\,(1-\cos^{(N-2)} (2\,\chi\, t)=-{\rm Re}\,b
\end{array}
\end{equation}
 and the corresponding invariants $(I_4,\ I_{12},\ I_{14})$ are given by, 
\begin{eqnarray} 
\label{eq:invku}
I_{4}&=&\cos^{2(N-1)} (\chi\, t), \nonumber\\
I_{12}&=&\frac{I_{4}}{2}\,(1+\cos^{(N-2)} (2\chi\, t)), \  \nonumber\\
I_{14}&=&- 2\, I_4\,\cos^{2(N-2)}(\chi\ t)\, \sin^2 (\chi\, t)
\end{eqnarray}
Note that $I_{14}$ is manifestly negative, highlighting the non-local %%@
nature of the state. 

\subsection{State extracted from 1-d Ising chain of N-qubits}
A one  dimensional chain of $N$ qubits with nearest neighbour interaction %%@
is characterized by the Ising model Hamiltonian %%@
$H=\displaystyle\frac{\chi}{4}\, \displaystyle\sum _{i=1}^N\sigma _{i %%@
x}\sigma _{(i+1)x}$, where $\chi$ denotes the interaction strength.
The Hamiltonian evolution of $N$ qubits, taken in a {\em all-down} initial %%@
state, results in a two-qubit density matrix~\cite{Xi01} with the state %%@
parameters, 
\begin{equation}
\begin{array}{l}
\displaystyle{a=\frac{4\, (N-1)\, (1+\cos^2(\frac{\chi\, %%@
t}{2}))-\sin^2(\chi\, t)}{8\, (N-1)}},\cr
\displaystyle{b=-\frac{\sin (\chi\, t)\, (\sin (\chi\, t)+4\, i)}{8\, %%@
(N-1)}},\cr 
 \displaystyle{c=\frac{\sin^2 (\chi\, t)}{8\, (N-1)}}
\end{array}
\end{equation}
 and the local entanglement invariants are given by, 
\begin{eqnarray*} 
\label{eq:invising}
I_{4}&=&\cos^{4} \left(\frac{\chi\, t}{2}\right),\ \\ 
I_{12}&=&I_{4}\,\left(1- \frac{\sin^{2} (\chi\, t)}{2\, (N-1)}\right),\\ 
I_{14}&=&-\frac{ 2\, I_4\,\sin^{2}(\chi\ t)}{(N-1)^2}\, .
\end{eqnarray*} 
In this model the local invariant $I_{14}$ assumes negative value %%@
indicating that  the system is entangled. 
\section{\label{sec:level4} Conclusions}
We have shown that a subset $(I_1,\  I_2,\   I_4,\ I_{10},\ I_{12},\ %%@
I_{14}),$ of a more general set of 18 invariants proposed by %%@
Makhlin~\cite{Mak02}, 
is   sufficient to characterize the non-local properties of a symmetric %%@
two-qubit system. Invariants $I_{12},\ I_{14}$ and $I_{12}-I_4^2$ of   %%@
separable symmetric two-qubit states are shown to be non-negative. We have 
 proposed sufficient conditions for identifying entanglement in symmetric %%@
two-qubit states, when the qubits have a non-zero 
value for the average spin. Moreover these conditions on the invariants are 
shown to be necessary and sufficient for a class of symmetric two qubit %%@
states. We have calculated the invariants of some physical examples of %%@
two-qubit states picked from $N$ qubit states like, 
 (i) symmetric Dicke state, (ii) state generated by one-axis twisting %%@
Hamiltonian and (iii) 1-d Ising model of $N$ qubits. We have 
explicitly  demonstrated  the non-separability of such states through our %%@
characterization in terms of  local invariants.


\begin{thebibliography}{1}
\bibitem{footnote1}The squared collective angular momentum operator %%@
satisfies the condition $S_x^2+ S_y^2+ %%@
S_z^2=S(S+1)=\frac{N}{2}(\frac{N}{2}+1)$ for symmetric states.
More explicitly we have, %%@
$\frac{1}{4}\langle(\vec{\sigma}_1+\vec{\sigma}_2)^2\rangle=
\frac{N}{2}(\frac{N}{2}+1)=2$, which in turn leads to condition ${\rm Tr} %%@
(T)=1$ on the correlation matrix of the two-qubit symmetric states.
\bibitem{footnote2} Note that if the product $s_1\,s_2\,s_3$ is positive, %%@
and say, 
 $(+, +, +)$ are the signs of each components,  identical local rotations %%@
about the  axes $i=1,\ 2,\ 3$ on the qubits,  
by an angle $180^\circ$ affect only  the signs of the 
components and lead  to the alternative possibilities  $(+, -, -)$, $(-, +, %%@
-)$ or  $(-, -, +),$  all of which have positive sign for the product 
$s_1\,s_2\,s_3$. Similarly, if the product $s_1\,s_2\,s_3$ is negative, and %%@
say, 
has $(-, -, -)$ signs for each components,  these identical local rotations %%@
through an angle $180^\circ$   change the signs of the 
components to  $(-, +, +)$, $(+, -, +)$ or  $(+, +, -),$  all of which have %%@
negative  sign for the product 
$s_1\,s_2\,s_3$ i.e., the absolute sign of the product  $s_1\,s_2\,s_3$ is %%@
a local invariant.
\end{thebibliography}
\end{document}